\documentclass[aps,prl,10pt,twocolumn,superscriptaddress,showpacs,amsmath,amssymb,floatfix,citeautoscript,longbibliography]{revtex4-1}
\usepackage{graphicx}
\usepackage[pdftex,bookmarks=false]{hyperref}

\begin{document}

\title{Stripe order from the perspective of the Hubbard model}

\author{Edwin W. Huang}
\affiliation{Department of Physics, Stanford University, Stanford, California 94305, USA}
\affiliation{Stanford Institute for Materials and Energy Sciences, SLAC National Accelerator Laboratory and Stanford University, Menlo Park, CA 94025, USA}
\author{Christian B. Mendl}
\affiliation{Stanford Institute for Materials and Energy Sciences, SLAC National Accelerator Laboratory and Stanford University, Menlo Park, CA 94025, USA}
\author{Hong-Chen Jiang}
\affiliation{Stanford Institute for Materials and Energy Sciences, SLAC National Accelerator Laboratory and Stanford University, Menlo Park, CA 94025, USA}
\author{Brian Moritz}
\affiliation{Stanford Institute for Materials and Energy Sciences, SLAC National Accelerator Laboratory and Stanford University, Menlo Park, CA 94025, USA}
\affiliation{Department of Physics and Astrophysics, University of North Dakota, Grand Forks, ND 58202, USA}
\author{Thomas P. Devereaux}
\email{tpd@stanford.edu}
\affiliation{Stanford Institute for Materials and Energy Sciences, SLAC National Accelerator Laboratory and Stanford University, Menlo Park, CA 94025, USA}
\affiliation{Geballe Laboratory for Advanced Materials, Stanford University, Stanford, CA 94305, USA}

\date{\today}

\begin{abstract}
A microscopic understanding of the strongly correlated physics of the cuprates must account for the translational and rotational symmetry breaking that is present across all cuprate families, commonly in the form of stripes. Here we investigate emergence of stripes in the Hubbard model, a minimal model believed to be relevant to the cuprate superconductors, using determinant quantum Monte Carlo (DQMC) simulations at finite temperatures and density matrix renormalization group (DMRG) ground state calculations. By varying temperature, doping, and model parameters, we characterize the extent of stripes throughout the phase diagram of the Hubbard model. Our results show that including the often neglected next-nearest-neighbor hopping leads to the absence of spin incommensurability upon electron-doping and nearly half-filled stripes upon hole-doping. The similarities of these findings to experimental results on both electron and hole-doped cuprate families support a unified description across a large portion of the cuprate phase diagram.
\end{abstract}

\maketitle

{\bf Introduction:} The lack of an analytic solution to the Hubbard model in two-dimensions has led to development of various numerical methods to study its low temperature and ground state properties. Calculations to benchmark these techniques have revealed that different candidate ground states all lie close in energy \cite{Corboz2014,Zheng2017}, with small differences possibly associated with specific aspects of each method. Density matrix renormalization group, exact diagonalization/dynamical mean-field theory, constrained path auxiliary field Monte Carlo, infinite projected entagled-pair states, and density matrix embedding theory all find evidence for stripes \cite{White1998,White2003,Hager2005,Fleck2000,Chang2010,Corboz2014,Zheng2017}, having stronger amplitudes and longer correlation lengths than $d$-wave superconductivity. However, dynamical cluster approximation and cellular dynamical mean-field theory calculations have not shown evidence for stripes, instead finding a finite temperature transition into a $d$-wave superconductor \cite{MaierRMP2005,Maier2005,Kancharla2008,Sordi2012,Gull2012,Gull2013}. These seemingly different ground states with similar energies reflect a delicate balance, sensitive to the specific nuances and biases of each approach.

Numerically discerning energy differences to ascertain low temperature properties requires rigorous effort to eliminate biases, and techniques may or may not reveal true ground states if the treatments are variational. On the other hand, provided that fluctuating orders are appreciable, calculations at higher temperatures provide an alternative perspective and carry the benefit that shortened correlation lengths reduce finite size effects. Here we use determinant quantum Monte Carlo (DQMC), an exact finite temperature technique, for this purpose. Although the fermion sign problem sets a lower bound on the range of temperatures amenable to simulation, we show that fluctuating stripe order is nevertheless observable at accessible temperatures.

\begin{figure}
\centering
\includegraphics[width=0.8\columnwidth]{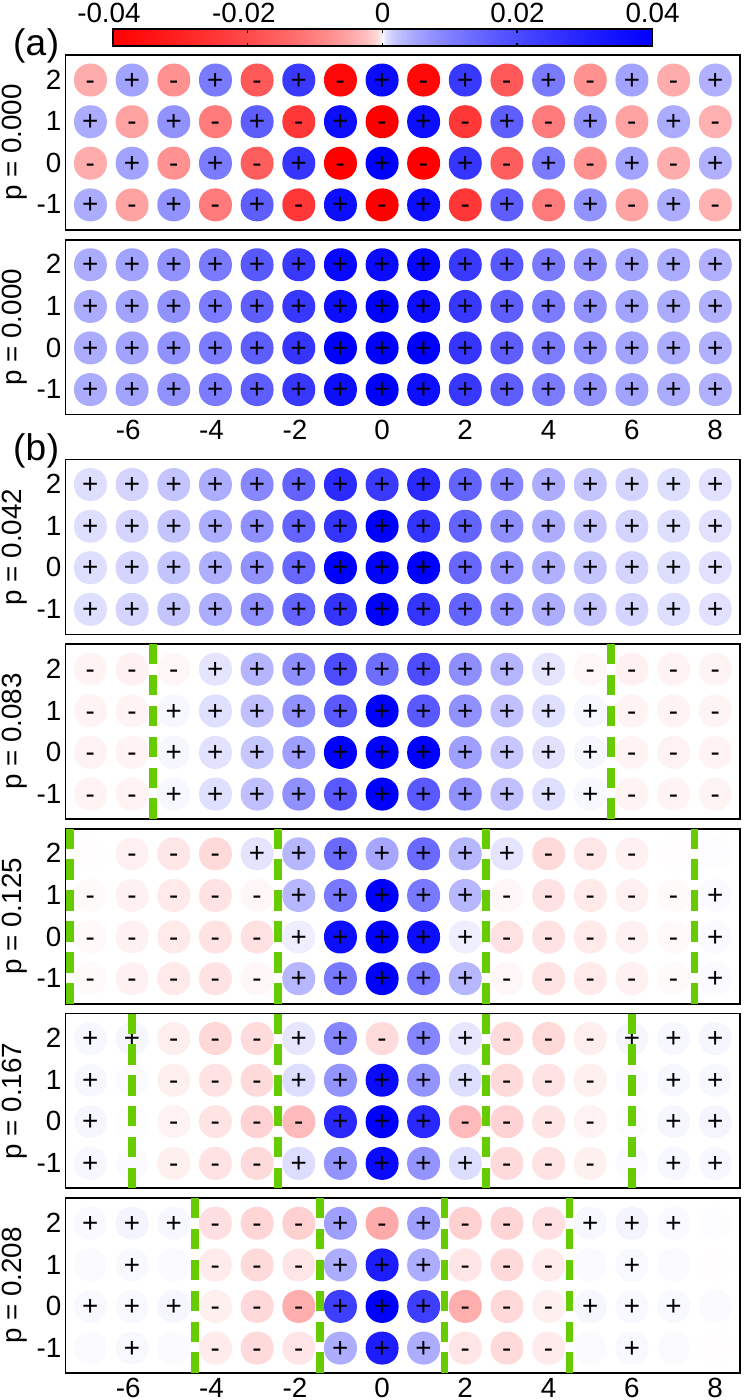}
\caption{(a) Spin correlation functions in the Hubbard model with parameters $U/t=6, t'/t=-0.25$ obtained by DQMC simulations at temperature $T/t= 0.22$. Top: Spin correlation functions at 0 doping. Bottom: staggered spin correlation functions, with signs flipped on every other site, of the same data. (b) Staggered spin correlation functions for various hole doping levels. Dashed green lines indicate approximate locations of antiphase domain walls. Correlations showing a + or - sign are nonzero by at least 2 standard errors.}
\label{fig:spin_corr}
\end{figure}

{\bf Results:} We first describe the doping dependence of spin correlations for the Hubbard model with interaction strength $U/t=6$ and next-nearest-neighbor hopping $t'/t=-0.25$, where $t$ is the nearest-neighboring hopping. Fig.~\ref{fig:spin_corr} displays the real space, equal-time spin-spin correlation functions obtained from finite temperature DQMC simulations on $16\times4$ clusters with periodic boundary conditions. At half-filling (Fig.~\ref{fig:spin_corr}a), antiferromagnetic spin correlations are evident from the checkerboard pattern, or equivalently from the uniform phase of the staggered spin-spin correlation functions (lower portion of Fig.~\ref{fig:spin_corr}a), defined with a $(\pi,\pi)$ phase factor that flips the sign of the correlation function on every other site. Hole-doping (Fig.~\ref{fig:spin_corr}b, $p = 0.042$) first induces a decrease in correlation length, followed by development of antiphase domains with increasing hole concentration. The size of each domain is inversely proportional to the hole doping level; and for $p \ge 0.125$, multiple sets of antiphase domain walls become visible for this cluster geometry and size. This behavior, qualitatively and quantitatively similar to previous findings for the three-band Hubbard model \cite{Huang2016}, demonstrates that stripe behavior at finite temperatures emerges in the Hubbard model through incommensurate spin correlations. To ensure that these findings are not artifacts of the anisotropic cluster geometry, we present and discuss results for a square geometry in Fig.~S1 of the Supplementary Materials.

Previous finite-temperature calculations of the Hubbard model failed to demonstrate spontaneous development of either spin or charge incommensurability, absent imposing inhomogeneity from external fields not part of the original model ({\it e.g.} see Refs.~\onlinecite{Maier2010} or \onlinecite{Mondaini2012}). In light of the results presented here, a necessary ingredient appears to be large enough clusters capable of supporting multiple stripe domains. While calculations utilizing small clusters ({\it e.g.} $N=8$) have been used to demonstrate antiferromagnetism, superconductivity, and pseudogap physics \cite{MaierRMP2005,Gull2013,Chen2016}, their inability to host incommensurate states leaves open important questions regarding the interplay of stripes with the aforementioned orders.

\begin{figure}
\centering
\includegraphics[width=0.8\columnwidth]{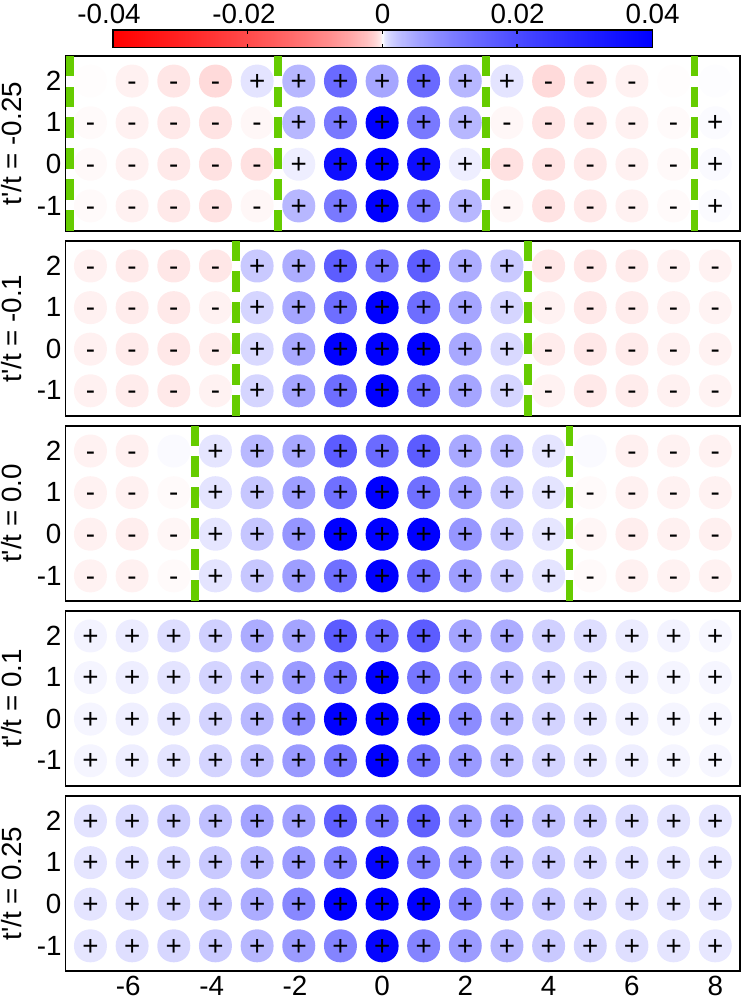}
\caption{Staggered spin correlation functions from DQMC at $p=0.125$ hole doping, $U/t=6$, and $T/t=0.22$ for various values of $t'$.}
\label{fig:tp}
\end{figure}

In contrast to these previous finite temperature findings, zero-temperature calculations, from a variety of methods, have indicated striped ground states in the Hubbard model. A recent comparison found close agreement in the ground state energies using four different techniques \cite{Zheng2017}, providing evidence for period-8 stripes in the ground state of the 1/8-hole-doped Hubbard model with only nearest-neighbor hopping ($t'=0$) and canonical interaction strength ($U/t=8$, the non-interacting bandwidth). Instead, our findings show stripes with a period $\sim 5$ for 1/8 hole-doping (Fig.~\ref{fig:spin_corr}b, $p=0.125$), in better agreement with experimental results on cuprates \cite{Yamada1998,Fujita2004,Enoki2013}.

To understand these differences, we study the impact of varying the model parameters, starting with the next-nearest-neighbor hopping $t'$, which induces a particle-hole asymmetry. For $t'=-0.1$ and $t'=0$, we find antiphase domain walls still present, but with an increased period of $\sim 8$ for 1/8 hole-doping (Fig.~\ref{fig:tp}), similar to the previously mentioned results of ground state calculations. In contrast, the period $\sim 5$ stripes from simulations using a negative value of $t'=-0.25$ correspond well to neutron scattering experiments where multiple hole-doped compounds show incommensurablity corresponding to period $4-5$ spin stripes at 1/8 hole-doping \cite{Enoki2013}. Also at 1/8 hole-doping, Fig.~\ref{fig:tp} shows the staggered spin-spin correlation function for $t'/t = 0.1$ and $t'/t = 0.25$, equivalent to 1/8 electron-doping for negative $t'$. In contrast to previous results, no antiphase domains are present and only antiferromagnetism is visible. This is additionally corroborated by our DMRG simulations in Fig.~S2 of the Supplementary Materials. As neutron scattering \cite{Yamada2003,Armitage2010} on electron-doped compounds similarly finds only commensurate antiferromagnetic excitations at low energy, our simulations show that a negative value of $t'$ that properly captures the cuprates' Fermi surface topology also correctly describes the spin behavior in both directions of doping.

\begin{figure*}
\centering
\includegraphics[width=2\columnwidth]{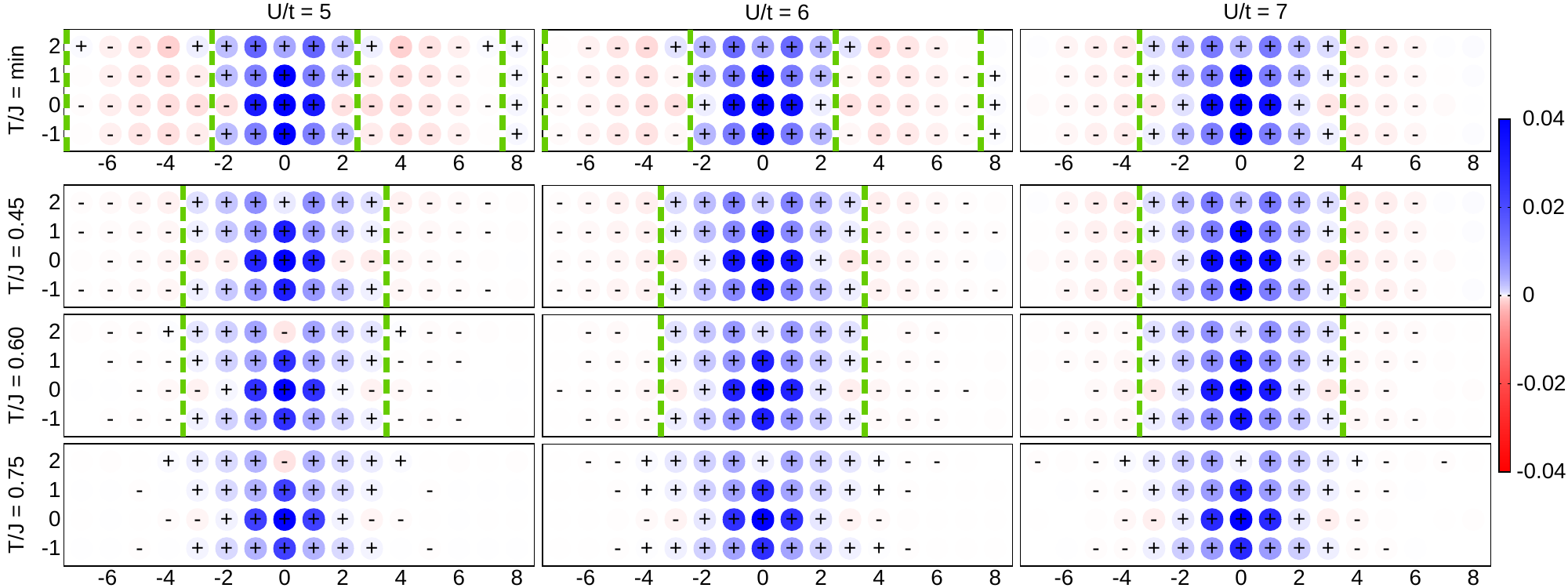}
\caption{Staggered spin correlation functions from DQMC at $p=0.125$ hole doping and $t'/t = -0.25$ for various values of $U$ and $T$. $J = 4 t^2/U$ is the leading order approximation of the exchange coupling constant. In the top row, the temperature is set to $T/t=0.20$ ($T/J = 0.25$), $T/t = 0.22$ ($T/J = 0.33$), and $T/t=0.26$ ($T/J = 0.45$) for $U/t = 5, 6, 7$ respectively.}
\label{fig:ut}
\end{figure*}

Variations in the interaction strength $U$ make little direct impact on the presence or periodicity of stripes. We first consider results for the lowest temperature accessible to simulation. For $U/t=5$, at a temperature of $T/t = 0.20$, we again find period-5 stripes at 1/8 hole-doping (Fig.~\ref{fig:ut}, top left). Increasing to $U/t=7$ (Fig.~\ref{fig:ut}, top right), the worsened sign problem constrains the lowest accessible temperature to $T/t = 0.26$. Here, the stripes instead have an increased period of $\sim 7$. We attribute this to the change in temperature: for the same ratio $T/J=0.45$ of temperature to exchange coupling, similar period $\sim 7$ domains are present for $U/t=5$ and $U/t=6$ (Fig.~\ref{fig:ut}, second row). At $U/t=8$, the sign problem is too severe to achieve temperatures of $T/J=0.45$, but at accessible temperatures we find no indication of different behavior. The similarities between these results for the same $T/J$ imply a marginal role of the value of $U$, at least in the range of explored values. 

Generally, with increasing temperatures we find slight increases in stripe period and substantial reduction in correlation length (Fig.~\ref{fig:ut}). This is consistent with neutron scattering data on LBCO, where spin incommensurability (inversely proportional to the period) decreases with increasing temperature \cite{Fujita2004}. In our data, reduced correlation lengths at higher temperatures make it increasingly difficult to see $\pi$-phase shifted domains in correlation functions. At the temperature $T/J=0.75$, nearly all correlations expected from the nearest $\pi$-phase shifted domain have magnitude within the simulations' sampling error. However, since sampling error is dependent on the length of the Monte Carlo simulation, the temperature that roughly delineates where stripes are visible should not necessarily be interpreted as an onset temperature.

The temperatures of our simulations correspond to that of the strange metal regime in cuprate phase diagram. Recent hydrodynamic studies have proposed fluctuating charge or spin density waves as a source of bad metallic behavior \cite{Delacretaz2016,Delacretaz2017}. Future work via similar finite temperature quantum Monte Carlo simulations could address these potential connections between fluctuating stripes and possible signatures of strange metal physics.

\begin{figure}
\centering
\includegraphics[width=1\columnwidth]{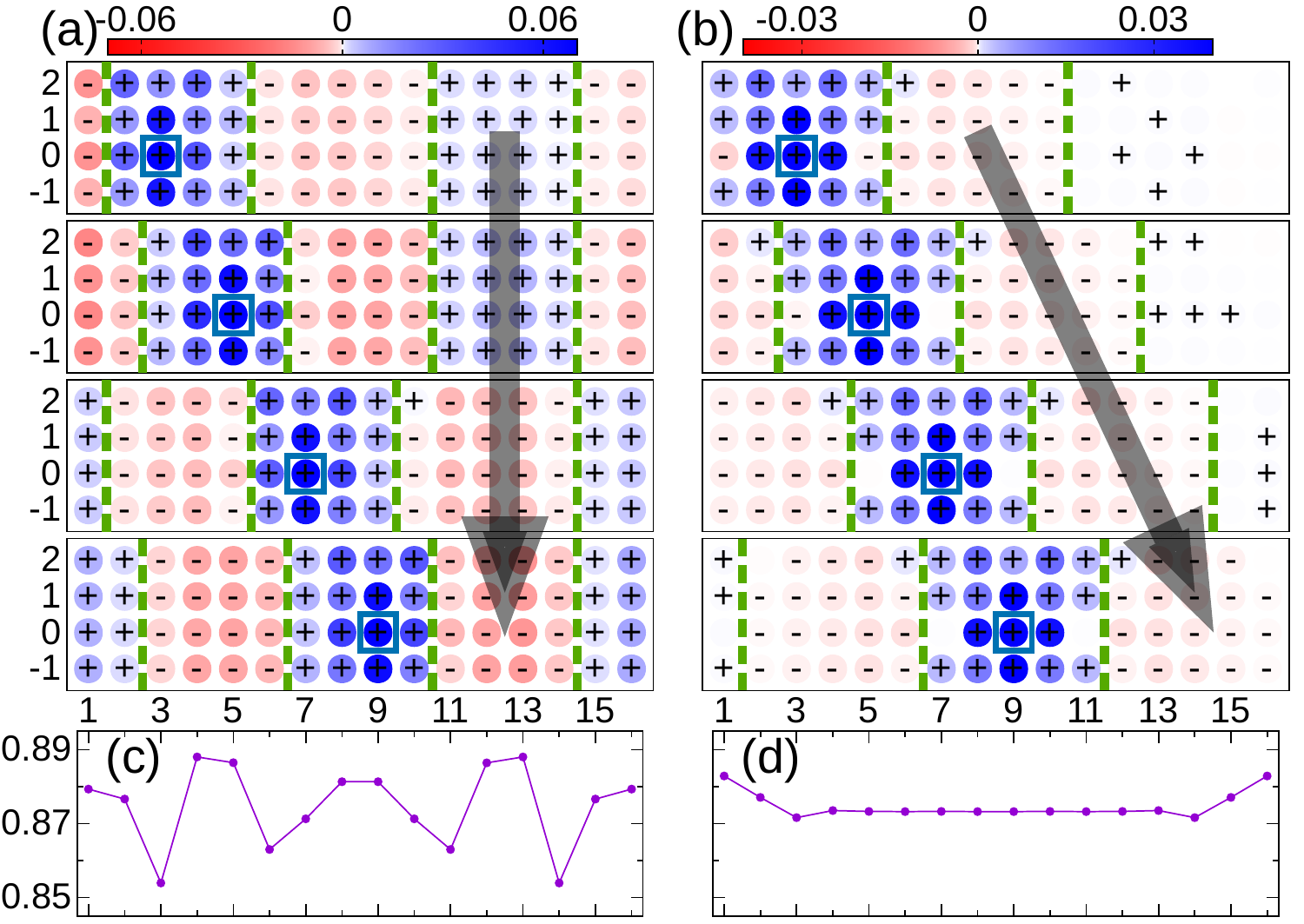}
\caption{(a) and (b) Staggered spin correlation functions with open left and right boundaries and periodic vertical boundaries, directly comparing (a) zero-temperature DMRG results with (b) DQMC calculations at finite temperature ($T/t=0.22$) using identical Hubbard model parameters ($U/t=6, t'/t=-0.25$). Boxes indicate reference site $\mathbf i$ and axes indicate the position of site $\mathbf j$ of the correlation function $\langle S_z(\mathbf i) S_z(\mathbf j) \rangle$. Correlations showing a + or - sign are nonzero by at least 2 standard errors. (c) and (d) Electron density profiles for DMRG and DQMC as functions of the horizontal position, averaged over equivalent vertical sites. Standard errors in (d) are $\sim 2.5 \times 10^{-5}$.}
\label{fig:cbc}
\end{figure}

\begin{figure*}
\centering
\includegraphics[width=2\columnwidth]{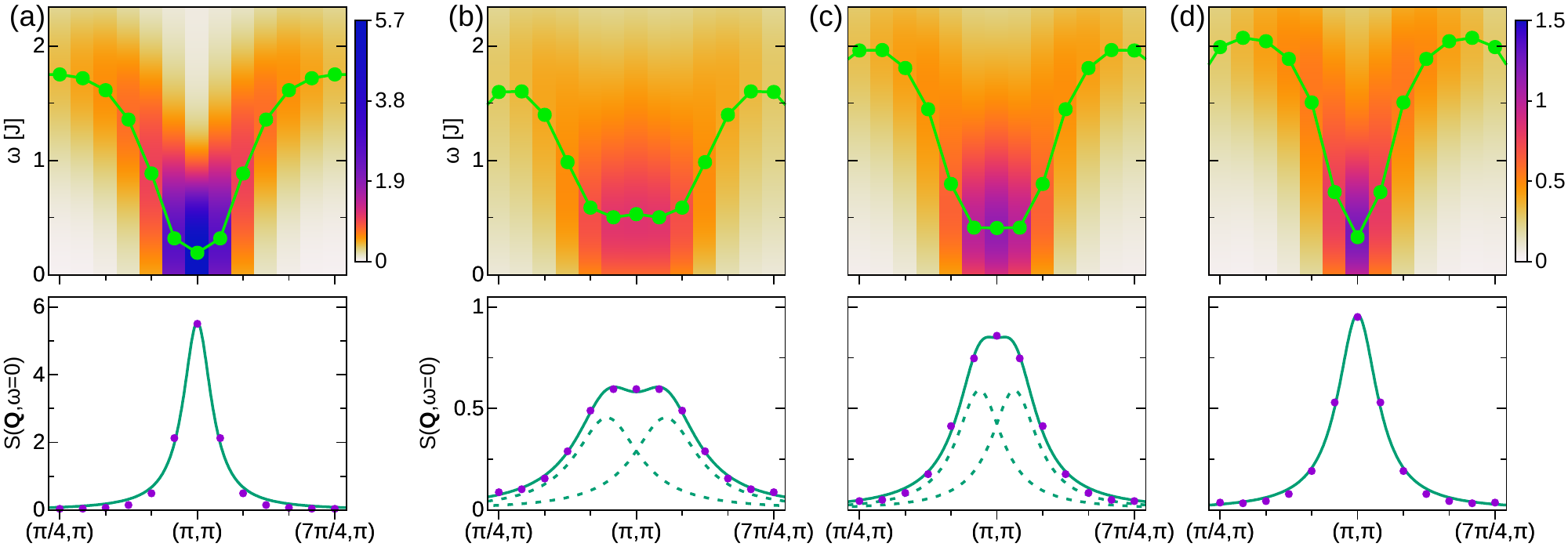}
\caption{Dynamical spin structure factor $S(\mathbf Q, \omega)$ along $Q_y = \pi$ calculated from the DQMC simulations of Figs.~\ref{fig:spin_corr} and \ref{fig:tp}. Common parameters for (a-d) are $U/t=6$ and $T/t=0.22$. The doping and next-nearest-neighbor hopping are (a) $p=0$, $t'/t=-0.25$, (b) $p=0.125$, $t'/t=-0.25$, (c) $p=0.125$, $t'=0$, and (d) $p=0.125$, $t'/t=0.25$. $J = 4t^2/U \approx 0.67 t$ is the leading order approximation of the exchange coupling constant. Top: False color maps of $S(\mathbf Q, \omega)$. Green dots are peak positions at each wavevector. Bottom: $S(\mathbf Q, \omega=0)$. Lines are fits to either single (a, d) or double (b, c) Lorentzian peaks.}
\label{fig:sqw}
\end{figure*}

We additionally compare our finite-temperature DQMC results to zero-temperature DMRG calculations\cite{White1992} for the same model parameters and lattice geometry with cylindrical boundary conditions. Figure~\ref{fig:cbc} displays the spin-spin correlation function and density profile from both techniques. Each shows anti-phase domains characteristic of spin stripes (Fig.~\ref{fig:cbc}a, b), with period-4 stripes at $T=0$ and period-5 stripes at $T/t=0.22$. The correlations at zero temperature (Fig.~\ref{fig:cbc}a) show immobile anti-phase domain walls that are pinned by the open boundaries. At finite temperatures, we observe short-ranged and mobile antiphase domain walls, as evidenced by their tendency to follow the reference spin position on the cluster, as expected from fluctuating stripes (Fig.~\ref{fig:cbc}b). This contrasting behavior is evident in the density profiles as well (Fig.~\ref{fig:cbc}c, d), where zero-temperature DMRG results reveal a static charge density wave, with peaks and troughs in the electron occupation coinciding with the antiphase domains, in agreement with the long-established picture of stripes known from the earliest mean-field studies \cite{Machida1989, Zaanen1989}. The finite temperature DQMC results instead show only minor modulation of the occupation due to boundary effects, without any indication of static charge order.

Finally, to understand the dynamical properties associated with spin stripes, we use the maximum entropy method (MEM) for analytic continuation \cite{Jarrell1996, Macridin2004} to extract the dynamical spin structure factor $S(\mathbf Q, \omega)$ from the finite temperature, unequal-time spin-spin correlation function obtained using DQMC. At zero doping (Fig.~\ref{fig:sqw}a), the dispersion is conical around $(\pi, \pi)$, as expected for antiferromagnetism. Upon hole doping to $p=0.125$ (Fig.~\ref{fig:sqw}b), the most drastic change is hardening and loss of spectral weight at $(\pi, \pi)$, as expected when antiferromagnetism is no longer dominant. While the spin excitations at the closest wavevectors accessible to our cluster $(7\pi/8, \pi)$ and $(9\pi/8, \pi)$ exhibit similar behavior, the excitations at $(3\pi/4, \pi)$ and $(5\pi/4, \pi)$ instead soften. This change is also reflected in zero-energy spin structure factor, which shows two peaks (Fig.~\ref{fig:sqw}b, bottom) split from $(\pi, \pi)$. In neutron scattering experiments on hole-doped cuprates, low energy incommensurate peaks at these wavevectors near $(\pi, \pi)$ are found and interpreted as evidence for stripes \cite{Tranquada1995,Tranquada2004,Hayden2004}. Although our data do not resolve sharp peaks as in neutron scattering data, which are taken at far lower temperatures, the agreement with the real-space data of Fig.~\ref{fig:spin_corr} strongly supports that doping induces not only weakened antiferromagnetism but also incommensurate spin correlations.

As in Fig.~\ref{fig:tp}, we study the effect of $t'$ by fixing the doping at $p=0.125$ and varying $t'/t$ from $-0.25$ (Fig.~\ref{fig:sqw}b) to $0$ (Fig.~\ref{fig:sqw}c) and $0.25$ (Fig.~\ref{fig:sqw}d). At $t'/t=0$, similar behavior is present as in $t'/t=-0.25$, but with a significantly smaller separation of the two low energy peaks. This is consistent with our real-space data in Fig.~\ref{fig:tp} that shows increased stripe period for $t'/t=0$. In contrast, for positive $t'/t=0.25$, low energy spectral weight is evidently centered at $(\pi, \pi)$, in agreement our previous real-space results in Fig.~\ref{fig:tp} that spin correlations are predominantly antiferromagnetic for $t'/t=0.25$. Subject to the limited momentum resolution of our calculation, the low energy behavior seen here is reminiscent of the commensurate excitations at $(\pi, \pi)$ seen in electron-doped cuprates \cite{Yamada2003,Armitage2010}.

{\bf Discussion:} We have presented the first unbiased study demonstrating fluctuating spin stripes in the Hubbard model at finite temperatures. Without applying external fields, we observe incommensurate spin correlations for a wide range of hole-doping at temperatures below roughly $T/J < 0.7$. We study the stripes' dependence on kinetic frustration ($t'$), finding that a value appropriate for cuprates ($t'/t=-0.25$) captures the experimentally observed doping asymmetry in the spin incommensurability. This result provides a simple resolution, without non-local Coulomb interactions, for the mismatch between experimental data on cuprates and ground-state calculations of the $t'=0$ Hubbard model.

The role of kinetic frustration for stripes may be understood from a strong-coupling perspective where magnetic moments are localized. Consider first the Hubbard model with only nearest neighbor hopping. When doped, a stripe state is more favorable than a uniformly doped antiferromagnetic state. In the former, virtual hopping of holes on antiphase domain walls preserves the local antiferromagnetism of stripe domains; in the latter, the motion of each individual hole leaves behind a trail of frustrated spins. The contribution of diagonal next-nearest neighbor hopping is the opposite: diagonal motion of individual holes in an antiferromagnetic background does not magnetically frustrate, but diagonal motion of holes in a stripe state breaks up the stripes and also frustrates the local antiferromagnetism. Thus, when diagonal motion becomes more energetically favorable (i.e. larger values of $t'/t$), the system tends more toward a uniform antiferromagnetic state. This is precisely what our simulations demonstrate in Fig.~\ref{fig:tp}; as $t'/t$ is adjusted from $-0.25$ to $+0.25$, the stripes grow in period until only uniform antiferromagnetism is present.

In our DMRG calculations, we find interlocked spin and charge stripes in the ground state of the simulated cluster. In contrast, for DQMC calculations using the same parameters and cluster we see instead only fluctuating and unpinned spin stripes down to the lowest accessible temperatures. Taken together, these results suggest that in the Hubbard model, static charge order sets in at temperatures beyond the scope of DQMC.

While incommensurate spin and charge density waves are present among multiple hole-doped cuprate families, interlocked stripes have been seen only in La-based compounds. In other compounds, while the charge ordering has wavevector (and hence periodicity) in the same range as that of La-based cuprates and is similarly pronounced around 1/8-doping, the doping dependence differs \cite{Comin2016}. In La-based cuprates, both charge ordering wavevectors and spin incommensurability (periodicities) increase (decrease) with doping. For other compounds, and possibly LBCO at higher temperatures \cite{Miao2017}, the charge ordering wavevector decreases with doping while the spin incommensurability, in observed cases, increases \cite{Enoki2013}, indicating decoupling of charge and spin ordering. Microscopic model calculations, such as the DMRG results presented here, demonstrate behavior similar to that in La-based cuprates. To date, the diverse and materials-specific behavior of charge ordering in the cuprates has not been captured in such calculations.

The growth of modern computing power combined with recent developments in computational techniques has allowed for significant progress toward understanding static properties of the Hubbard model \cite{Zheng2017}. However, to connect with experimental results, it has become necessary to determine and benchmark the dynamical properties. Here, our calculation of the dynamical spin structure factor captures both the falloff of antiferromagnetism and the development of low energy incommensurate peaks indicative of stripes. As our spectra's resolution is ultimately limited by cluster size and temperature, both of which are severely constrained by the fermion sign problem, development of new techniques and algorithms capable of calculating dynamics in larger-scale simulations will allow for a closer comparison against data from neutron and X-ray scattering experiments on the cuprates.

\section{Methods}

{\bf Hubbard model.}
The Hubbard model Hamiltonian is
\begin{equation}
H = - \sum_{i j \sigma} t_{ij}\, c_{i\sigma}^\dagger c_{j\sigma}^{\phantom{\dagger}} + U \sum_i \hat{n}_{i\uparrow} \hat{n}_{i\downarrow} - \mu \sum_{i\sigma}\hat{n}_{i\sigma}
\end{equation}
where $c_{i\sigma}^{\dagger}$ ($c_{i\sigma}^{\phantom{\dagger}}$) creates (annihilates) an electron with spin $\sigma$ at site $i$; $\hat{n}_{i\sigma}^{\phantom{\dagger}} = c_{i\sigma}^{{\dagger}} c_{i\sigma}^{\phantom{\dagger}}$; the hopping amplitude $t_{ij}$ is equal to $t$ (the simulation energy scale) if $i$ and $j$ are nearest-neighbors and equal to $t'$ for next-nearest neighbors; $U$ is the on-site repulsive Coulomb interaction; and the chemical potential $\mu$ controls the doping level.

{\bf Determinant quantum Monte Carlo.}
We perform DQMC simulations on the Hubbard model \cite{Blankenbecler1981,White1989} with parameters $U/t$ and $t'/t$ and temperature $T/t$ as indicated in the corresponding text and figure captions. The chemical potential is tuned to achieve the desired doping level to within an accuracy of $O(10^{-4})$. The imaginary time interval $[0,\beta]$ is partitioned into steps of size $0.07 < \Delta\tau < 0.12$, resulting in a negligible Trotter error for our parameters.

To ensure numerical stability in computing the equal-time Green's functions, we use the prepivoting stratification algorithm as described in Ref.~\onlinecite{Tomas2012}, allowing up to 8 matrix multiplications before performing a QR decomposition. The unequal-time Green's functions are constructed using the Fast Selected Inversion algorithm described in Ref.~\onlinecite{Jiang2016}, with blocks corresponding to the product of matrices from 8 time steps.

Equal-time measurements are performed 20 times per full space-time sweep. As computing the unequal-time Green's function is of comparable computational cost as a Monte Carlo sweep, unequal-time quantities are measured every 4th sweep. For each Markov chain, we use 50000 sweeps for warmup and 1000000 sweeps for measurements. Between 2000 and 60000 independently seeded Markov chains are used for each parameter set. This large amount of data allowed for excellent statistics despite the severe fermion sign problem: standard errors in the plotted spin correlation functions are $O(10^{-6})$ while the average sign dipped as low as $\sim0.015$ for $p > 0.1$.

{\bf Density matrix renormalization group.}
Using the same model parameters, we perform the standard DMRG simulations \cite{White1992} with up to 26 sweeps and keep up to $m=10000$ DMRG block states with a typical truncation error of $2\times 10^{-6}$ per step. This leads to the excellent convergence for the results that we report here.

{\bf Equal-time spin-spin correlation function.}
The equal-time spin-spin correlation function is
\begin{equation}
\langle S_z(\mathbf i) S_z(\mathbf j) \rangle
\end{equation}
where $S_z(\mathbf i) = \frac{1}{2}\left(\hat{n}_{\mathbf i\uparrow} - \hat{n}_{\mathbf i\downarrow} \right)$ is the $z$-component of spin on site $\mathbf i$. To reduce statistical errors, we average correlations over pairs of sites equivalent by the translation and reflection symmetries of the cluster.

{\bf Maximum entropy method analytic continuation.}
The imaginary time susceptibility is obtained by Fourier transforming the imaginary time spin-spin correlation function $\langle T_\tau S_z(\mathbf i, \tau) S_z(\mathbf j, 0) \rangle$ measured in DQMC. It is related to the real frequency susceptibility by
\begin{equation}
\chi(\mathbf Q,\tau) = \int_0^\infty \frac{d\omega}{\pi} \frac{e^{-\tau\omega} + e^{-(\beta-\tau)\omega}}
{1 - e^{-\beta\omega}} {\rm Im} \chi(\mathbf Q,\omega)
\label{eq:chi}
\end{equation}
The dynamical spin structure factor is calculated by the fluctuation-dissipation theorem, which simplifies to
\begin{equation}
S(\mathbf Q,\omega) = \frac{{\rm Im} \chi(\mathbf Q,\omega)}{1-e^{-\beta\omega}}
\label{eq:flucdis}
\end{equation}
Since inverting Eq.~\ref{eq:chi} is numerically ill-posed, we use Maximum Entropy analytic continuation \cite{Jarrell1996} to extract ${\rm Im}\chi(\mathbf Q,\omega)$ from the DQMC data. The classic variant of MEM is used with Bryan's method for optimization. A Lorentzian centered at $2J$ with width $1J$ is chosen as a minimally informative model function with the expected high-frequency decay.

{\bf Error analysis.}
For our DQMC data, we use jackknife resampling to calculate standard errors. For the analytically continued spectra, we verify the repeatability and reliability of the MEM procedure by repeating the calculation on disjoint sets of data bins.

{\bf Data availability.}
The data that support the findings of this study are available from the corresponding author upon reasonable request.

\section{Acknowledgements}

We thank Douglas Scalapino for helpful discussions and Shenxiu Liu for preliminary investigations. This work was supported by the U.S.~Department of Energy (DOE), Office of Basic Energy Sciences, Division of Materials Sciences and Engineering, under Contract No.~DE-AC02-76SF00515. Computational work was performed on the Sherlock cluster at Stanford University and on resources of the National Energy Research Scientific Computing Center, supported by the U.S.~DOE under Contract No.~DE-AC02-05CH11231. C.B.M.~acknowledges support from the Alexander von Humboldt Foundation via a Feodor Lynen fellowship.

\section{Author Contributions}

E.W.H.~and C.B.M.~performed the DQMC simulations and MEM analytic continuation. H.C.J.~performed the DMRG simulations. E.W.H.~analyzed the data with assistance from B.M. T.P.D.~and B.M.~supervised the project. All authors discussed the results and participated in writing the manuscript.

\section{Competing Interests}

The authors declare no competing interests.

\bibliographystyle{apsrev4-1}
%

\pagebreak
\renewcommand{\thefigure}{S\arabic{figure}}
\setcounter{figure}{0}

\section{Supplementary materials}
\subsection{Square geometry simulations}
\begin{figure}[h]
\centering
\includegraphics[width=0.9333333\columnwidth]{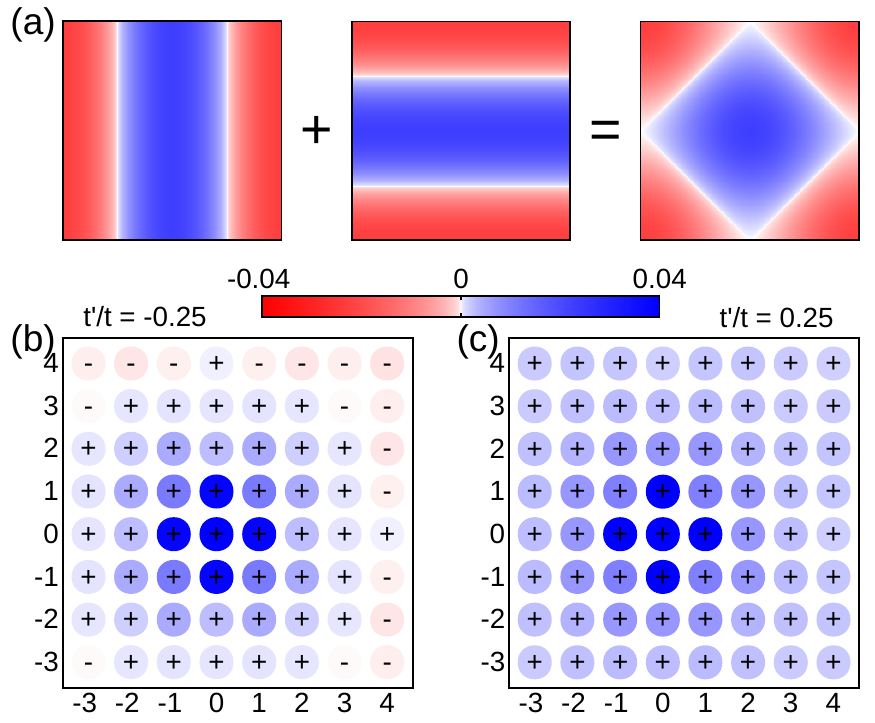}
\caption{(a) Illustration of expected staggered spin correlation functions for a superposition of horizontal and vertical stripes. (b) and (c) Staggered spin correlation functions calculated by DQMC for parameters $U/t=6, t'/t=\pm0.25, T/t=0.22, p=0.125$.}
\label{fig:sqr}
\end{figure}

To ensure that our results are robust against variations in cluster geometry, we perform DQMC calculations on a $8 \times 8$ square cluster. Our findings are summarized in Fig.~S1. Assuming the presence of the same stripes as demonstrated in the main text, but without breaking of $C_4$ rotational symmetry, the square cluster's spin correlations should appear as an equal superposition of correlations from horizontal and vertical stripes (Fig.~S1a). The predicted diamond pattern from such a superposition is observed in Fig.~S1b for the parameters $U/t=6, t'/t=-0.25, T/t=0.22, p=0.125$, which for the rectangular geometry we have observed period-5 stripes (Fig.~\ref{fig:spin_corr}). We note that the rough shape of the diamond is most likely related to the imperfect compatibility of period-5 stripes with the $8 \times 8$ period cluster. Unfortunately, the severity of the fermion sign problem ($\sim 0.01$ for the $8 \times 8$ clusters) prevents us from characterizing any larger clusters. Nevertheless, the qualitative behavior of the correlations and their distinction from the purely commensurate antiferromagnetism in Fig.~S1c ($t'/t=+0.25$) provide a picture of stripes consistent with our results for the rectangular cluster.

\subsection{Effect of varying $t'/t$ from DMRG}
\begin{figure}
\centering
\includegraphics[width=0.8\columnwidth]{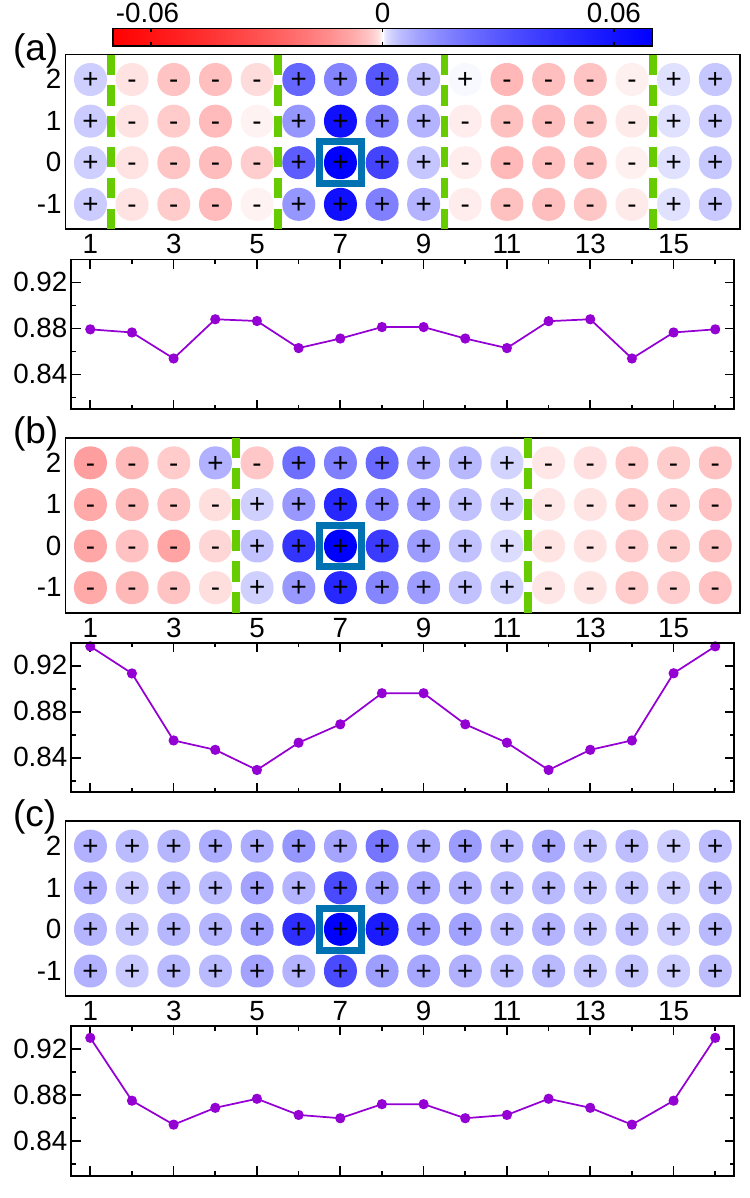}
\caption{(a-c) Staggered spin correlation functions (top) and charge density profiles (bottom) from DMRG simulations with parameters $U/t=6, T/t=0, p=0.125$ and at $t'/t = -0.25$ (a), $0.0$ (b), and $0.25$ (c).}
\label{fig:dmrgtp}
\end{figure}

In addition to our DQMC results in Fig.~\ref{fig:tp}, the dependence of stripe behavior on $t'$ can be studied by DMRG. Fig.~S2 displays the spin correlations and charge density profiles from our DMRG simulations for parameters $p=0.125, U/t=6$ and $t'/t=-0.25, 0.0, +0.25$. As in Fig.~\ref{fig:tp}, the spin stripes roughly double in period going from $t'/t=-0.25$ to $0.0$, and are absent for $t'/t=+0.25$. At $t'/t=-0.25$ and $0.0$, charge stripes are locked to the spin stripes, with maxima in the hole occupancy aligned to spin antiphase domain walls. For $t'/t=+0.25$, while the spin structure is purely antiferromagnetic, charge stripes with period 4 are present. We note that in electron-doped cuprates, recent X-ray scattering experiments have observed a signal at momentum transfer $q\sim(0.25, 0)$ in r.l.u., possibly arising from period 4 charge ordering \cite{daSilvaNeto2015,daSilvaNeto2016,Jang2017}.

\end{document}